\documentclass{PoS}
\usepackage{amsmath}
\usepackage{cancel}   % For S4b
\pdfoutput=1

\newcommand{\be}{\ensuremath{\beta} }
\newcommand{\Sb}{\ensuremath{\cancel{S^4}} }
\newcommand{\MSbar}{\ensuremath{\overline{\textrm{MS}} } }
\newcommand{\fig}[1]{Fig.~\ref{#1}}
\newcommand{\refcite}[1]{Ref.~\cite{#1}}
\newcommand{\secref}[1]{Section~\ref{#1}}
\title{Improved Lattice Renormalization Group Techniques}
\ShortTitle{Improved Lattice Renormalization Group Techniques}

\author{\speaker{Gregory Petropoulos}, Anqi Cheng, Anna Hasenfratz, David Schaich\footnote{Present address: Department of Physics, Syracuse University, Syracuse, NY 13244} \\
        Department of Physics, University of Colorado, Boulder, CO 80309 \\
        E-mail: \email{gregory.petropoulos@colorado.edu}}

\abstract{ % Draft complete
  We compute the bare step-scaling function $s_b$ for SU(3) lattice gauge theory with $N_f = 12$ massless fundamental fermions, using the non-perturbative Wilson-flow-optimized Monte Carlo Renormalization Group two-lattice matching technique.
  We use a short Wilson flow to approach the renormalized trajectory before beginning RG blocking steps.
  By optimizing the length of the Wilson flow, we are able to determine an $s_b$ corresponding to a unique discrete \be function, after a few blocking steps.
  We carry out this study using new ensembles of 12-flavor gauge configurations generated with exactly massless fermions, using volumes up to $32^4$.
  The results are consistent with the existence of an infrared fixed point (IRFP) for all investigated lattice volumes and number of blocking steps.
  We also compare different renormalization schemes, each of which indicates an IRFP at a slightly different value of the bare coupling, as expected for an IR-conformal theory.
}

\FullConference{31st International Symposium on Lattice Field Theory - LATTICE 2013 \\
                July 29 -- August 3, 2013 \\
                Mainz, Germany}
\begin{document}
%\section{Introduction} % Draft complete
For the past several years many lattice groups have been involved in studying strongly-coupled near-conformal gauge--fermion systems.
Some of these models may be candidates for new physics beyond the standard model, while others are simply interesting non-perturbative quantum field theories.
Because the dynamics of these lattice systems are unfamiliar, it is important to study them with several complementary techniques.
Not only does this allow consistency checks, it can also provide information about the most efficient and reliable methods to investigate near-conformal lattice theories.

Monte Carlo Renormalization Group (MCRG) two-lattice matching is one of several analysis tools that we are using to investigate SU(3) gauge theories with many massless fermion flavors.
This technique predicts the step-scaling function $s_b$ in the bare parameter space.
In a previous work~\cite{Petropoulos:2012mg} we proposed an improved MCRG method that exploits the Wilson flow to obtain a bare step-scaling function that corresponds to a unique discrete \be function.
We briefly review our Wilson-flow-optimized MCRG (WMCRG) procedure in Sections~\ref{sec:mcrg}--\ref{sec:wmcrg}.
It is important to note that we are investigating a potential infrared fixed point (IRFP) where the coupling is irrelevant: its running slows and eventually stops.
This is challenging to distinguish from a near-conformal system where the gauge coupling runs slowly but does not flow to an IRFP.
The observation of a backward flow that survives extrapolation to the infinite-volume limit could provide a clean signal.
In \secref{sec:results} we report WMCRG results for SU(3) gauge theory with $N_f = 12$ flavors of massless fermions in the fundamental representation.

This 12-flavor model has been studied by many groups, including Refs.~\cite{Appelquist:2009ty, Deuzeman:2009mh, Fodor:2011tu, Appelquist:2011dp, Hasenfratz:2011xn, DeGrand:2011cu, Cheng:2011ic, Jin:2012dw, Lin:2012iw, Aoki:2012eq, Fodor:2012uw, Fodor:2012et, Itou:2012qn, Cheng:2013eu, Aoki:2013pca, Hasenfratz:2013uha, Hasenfratz:2013eka, Cheng:2013bca}.
Using new ensembles of 12-flavor gauge configurations generated with exactly massless fermions, our improved WMCRG technique predicts a conformal IRFP where the step-scaling function vanishes.
As with every method, it is essential to study the systematic effects.
For WMCRG the most important systematic effects are due to the finite volume and limited number of blocking steps.
While we are not able to carry out a rigorous infinite-volume extrapolation, the observed zero of the bare step-scaling function is present for all investigated lattice volumes and renormalization schemes, and agrees with the earlier MCRG results of \refcite{Hasenfratz:2011xn}.
The results of our complementary $N_f = 12$ investigations of finite-temperature phase transitions~\cite{Schaich:2012fr, Hasenfratz:2013uha}, the Dirac eigenmode number~\cite{Cheng:2013eu, Cheng:2013bca}, and finite-size scaling~\cite{Hasenfratz:2013eka} are also consistent with the existence of an infrared fixed point and IR conformality.
% ------------------------------------------------------------------

% ------------------------------------------------------------------
\section{\label{sec:mcrg}Monte Carlo Renormalization Group} % Draft complete
MCRG techniques probe lattice field theories by applying RG blocking transformations that integrate out high-momentum (short-distance) modes, moving the system in the infinite-dimensional space of lattice-action couplings.
In an IR-conformal system on the $m = 0$ critical surface, a renormalized trajectory runs from the perturbative gaussian FP (where the gauge coupling \be is a relevant operator) to the IRFP (where \be is irrelevant).
Because the locations of these fixed points in the action-space depend on the renormalization scheme, each scheme corresponds to a different renormalized trajectory.
The RG flow produced by the blocking steps moves the system towards and along the renormalized trajectory, from the perturbative FP to the infrared fixed point.
At stronger couplings, where we would na\"ively expect backward flow, there might be no ultraviolet FP to drive the RG flow along a renormalized trajectory.
Except in the immediate vicinity of the IRFP, every method that attempts to determine the strong-coupling flow of the gauge coupling (including MCRG two-lattice matching) might then become meaningless.

We determine the bare step-scaling function $s_b(\be_1)$ by matching the lattice actions $S(\be_1, n_b)$ and $S(\be_2, n_{b - 1})$ for systems with bare couplings $\{\be_1, \be_2\}$ after $\{n_b, n_{b - 1}\}$ blocking steps: $s_b(\be_1) \equiv \lim_{n_b \to \infty} \be_1 - \be_2$~\cite{Petropoulos:2012mg}.
When the lattice actions are identical, all observables are identical.
We use the plaquette, the three six-link loops and a planar eight-link loop to perform this matching.
Using short-distance gauge observables allows us to carry out more blocking steps, down to small $2^4$ or $3^4$ lattices.
We minimize finite-volume effects by comparing observables measured on the same blocked volume~\cite{Hasenfratz:2011xn}.
We perform the matching independently for each observable, fitting the data as a cubic function of \be to smoothly interpolate between investigated values of the gauge coupling.

Our finite lattices only allow a few blocking steps, so we must optimize the procedure to reach the renormalized trajectory in as few steps as possible.
In practice, we optimize by tuning some parameter so that consecutive RG blocking steps yield the same $\be_1 - \be_2$, which we identify as $s_b(\be_1)$.
Traditional optimization tunes the RG blocking transformation at each coupling separately, resulting in a different renormalization scheme at each bare coupling $\be_1$: the $s_b$ we obtain is a composite of many different discrete \be functions.
The Wilson flow provides a parameter that we can tune without changing the scheme.
% ------------------------------------------------------------------

% ------------------------------------------------------------------
\section{\label{sec:wmcrg}Wilson-flow-optimized MCRG} % Draft complete
\begin{figure}[th]
  \centering
  \includegraphics[height=3 in]{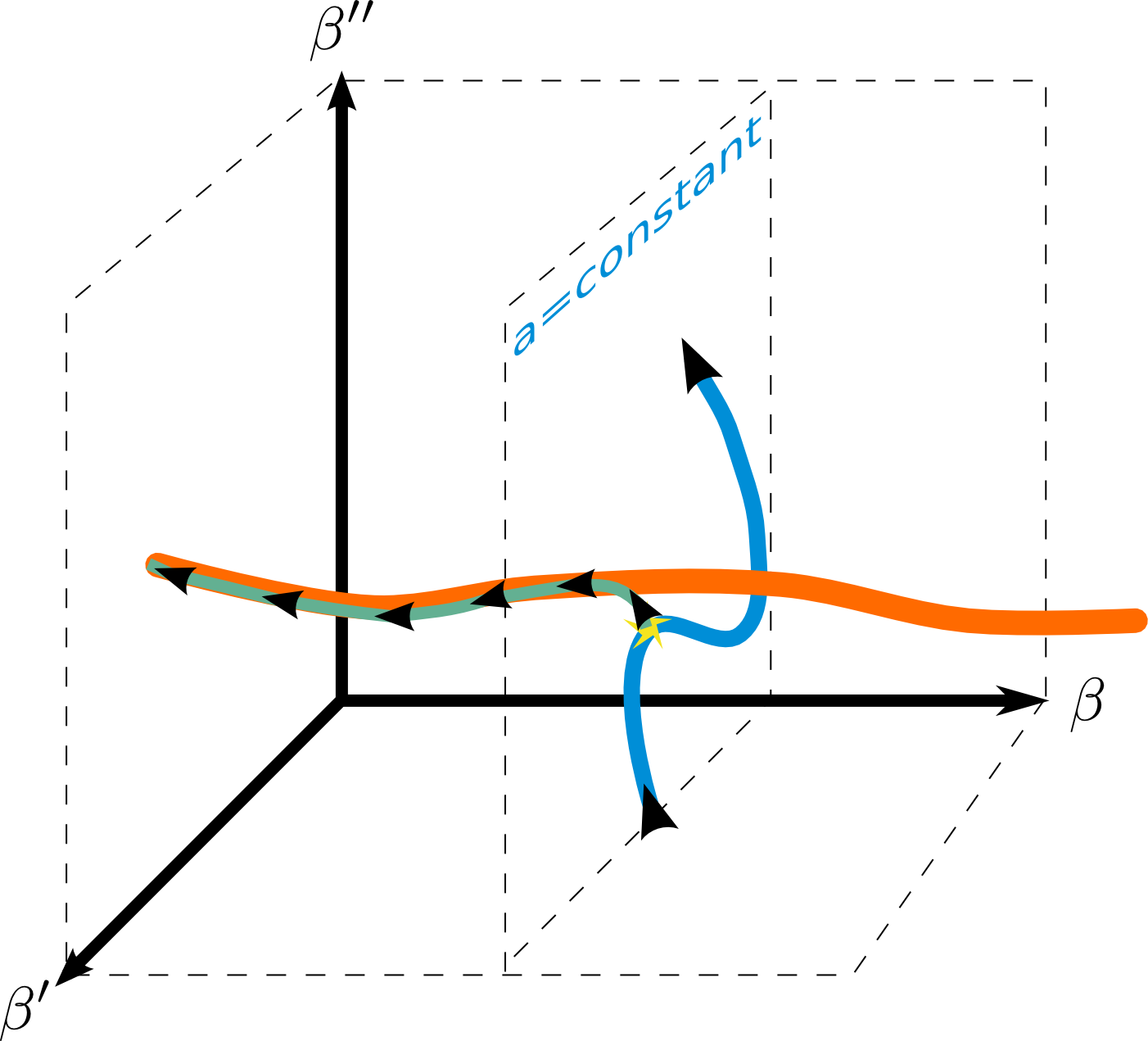}
  \caption{The Wilson flow (blue) moves systems on a surface of constant lattice scale $a$ (normal to the orange renormalized trajectory) in the infinite-dimensional coupling space.  Wilson-flow-optimized MCRG tunes the flow time to bring the system close to the renormalized trajectory (yellow star), so that MCRG blocking (green) quickly reaches the renormalized trajectory.}
  \label{fig:wflow_opt}
\end{figure}

The Wilson flow is a continuous smearing transformation~\cite{Narayanan:2006rf} that removes UV fluctuations without changing the lattice scale, as shown in \fig{fig:wflow_opt}.
In perturbation theory it is related to the \MSbar running coupling~\cite{Luscher:2010iy}, and can be used to compute a renormalized step-scaling function~\cite{Fodor:2012td, Fodor:2012qh}.

In this work we use the Wilson flow to optimize MCRG two-lattice matching with a fixed RG blocking transformation (renormalization scheme).
The Wilson flow continuously moves the system on a surface of constant lattice scale in the infinite-dimensional space of lattice-action couplings.
We tune the flow time to bring the system as close as possible to the renormalized trajectory.
After running the optimal amount of Wilson flow on the unblocked lattices, we then carry out the MCRG two-lattice matching.
Because the renormalization scheme is fixed, we obtain a bare step-scaling function that corresponds to a unique discrete \be function.
% ------------------------------------------------------------------

% ------------------------------------------------------------------
\section{\label{sec:results}Results for 12 Flavors} % Draft complete
\begin{figure}[ht]
  \centering
  \includegraphics[height=3 in]{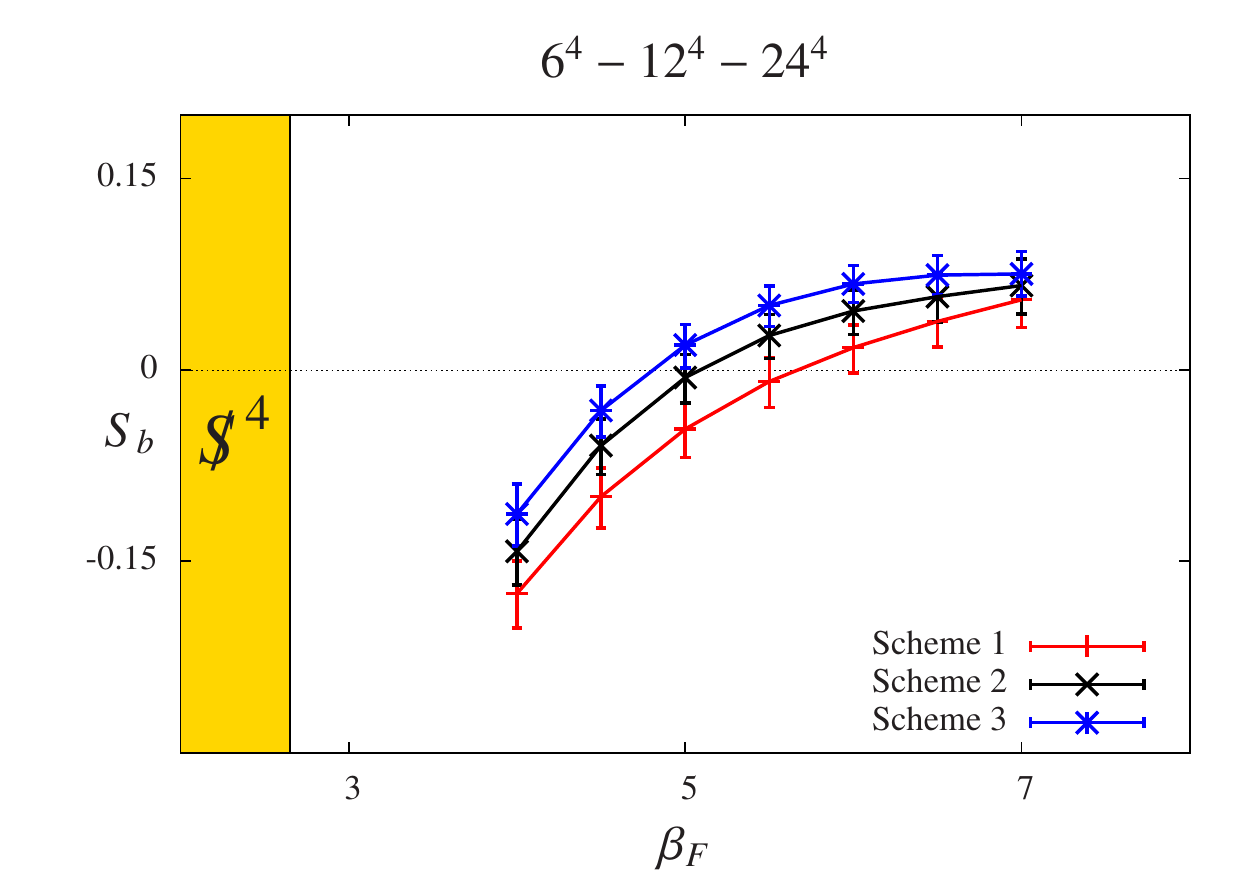}
  \caption{The bare step-scaling function $s_b$ predicted by three-lattice matching with $6^4$, $12^4$ and $24^4$ lattices blocked down to $3^4$, comparing three different renormalization schemes.  The error bars come from the standard deviation of predictions using the different observables discussed in \protect\secref{sec:mcrg}.}
  \label{fig:24to3}
\end{figure}

\begin{figure}[ht]
  \centering
  \includegraphics[height=3 in]{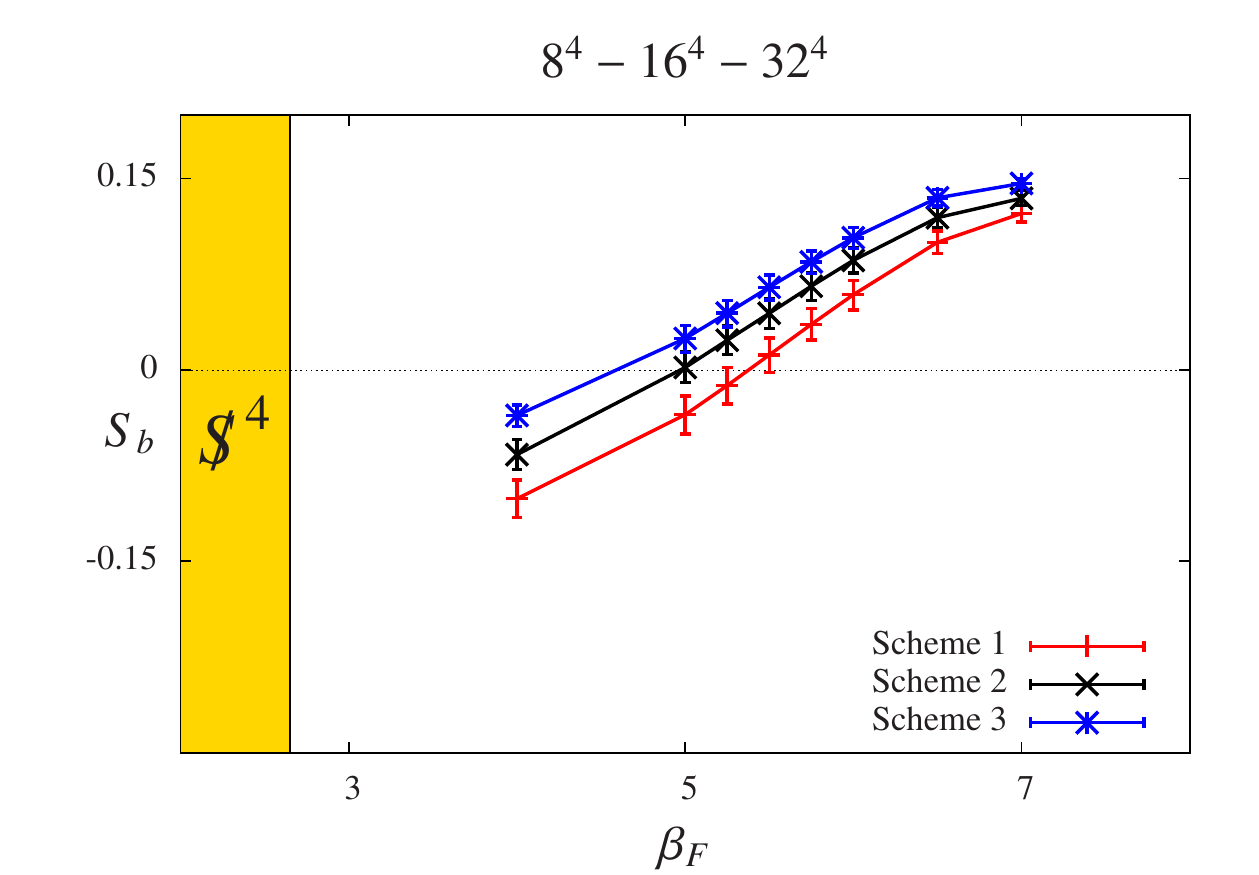}
  \caption{As in \protect\fig{fig:24to3}, the bare step-scaling function $s_b$ for three different renormalization schemes from three-lattice matching, now using $8^4$, $16^4$ and $32^4$ lattices blocked down to $4^4$.}
  \label{fig:32to4}
\end{figure}

Our WMCRG results for the 12-flavor system are obtained on gauge configurations generated with exactly massless fermions.
Our lattice action uses nHYP-smeared staggered fermions as described in \refcite{Cheng:2011ic}, and to run with $m = 0$ we employ anti-periodic boundary conditions in all four directions.
All of our analyses are carried out at couplings weak enough to avoid the unusual strong-coupling ``$\Sb$'' phase discussed by Refs.~\cite{Cheng:2011ic, Hasenfratz:2013uha}.

We perform three-lattice matching with volumes $6^4$--$12^4$--$24^4$ and $8^4$--$16^4$--$32^4$.
Three-lattice matching is based on two sequential two-lattice matching steps, to minimize finite-volume effects~\cite{Hasenfratz:2011xn}.
Both two-lattice matching steps are carried out on the same final volume $V_f$.
We denote the number of blocking steps on the largest volume by $n_b$, and tune the length of the initial Wilson flow by requiring that the last two blocking steps predict the same step-scaling function.
Using the $8^4$--$16^4$--$32^4$ data we determine the bare step-scaling function for $n_b = 3$ and $V_f = 4^4$ as well as $n_b = 4$ and $V_f = 2^4$, while the $6^4$--$12^4$--$24^4$ data set is blocked to a final volume $V_f = 3^4$ ($n_b = 3$).
This allows us to explore the effects of both the final volume and the number of blocking steps.
We investigate three renormalization schemes by changing the HYP smearing parameters in our blocking transformation~\cite{Petropoulos:2012mg}: scheme~1 uses smearing parameters (0.6, 0.2, 0.2), scheme~2 uses (0.6, 0.3, 0.2) and scheme~3 uses (0.65, 0.3, 0.2).

\begin{figure}[th]
  \centering
  \includegraphics[height=3 in]{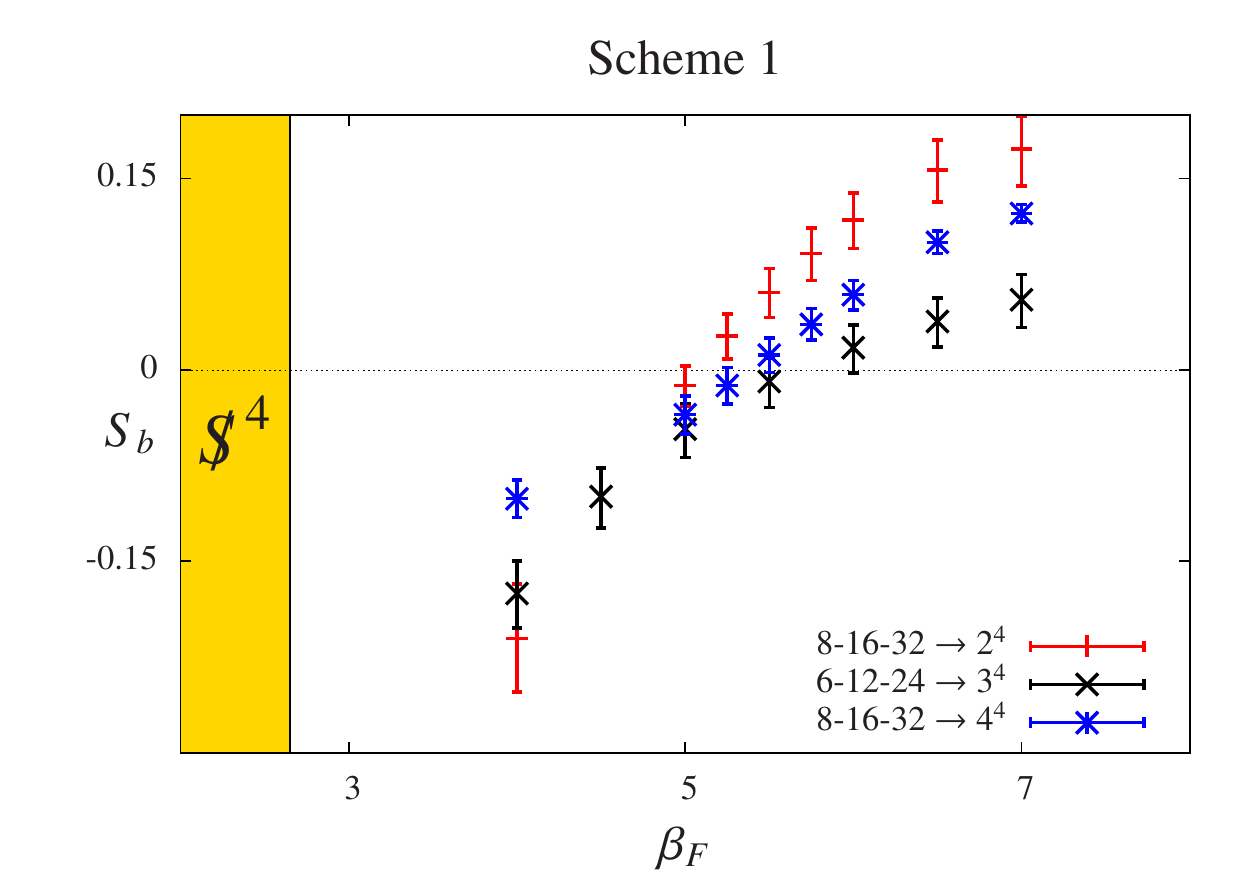}
  \caption{The bare step-scaling function $s_b$ for scheme~1, comparing three-lattice matching using different volumes: $6^4$, $12^4$ and $24^4$ lattices blocked down to $3^4$ (black $\times$s) as well as $8^4$, $16^4$ and $32^4$ lattices blocked down to $4^4$ (blue bursts) and $2^4$ (red crosses).}
  \label{fig:scheme1}
\end{figure}

Figs.~\ref{fig:24to3}, \ref{fig:32to4} and \ref{fig:scheme1} present representative results for 12 flavors.
All of the bare step-scaling functions clearly show $s_b = 0$, signalling an infrared fixed point, for every $n_b$, $V_f$ and renormalization scheme.
Appropriately for an IR-conformal system, the location of the fixed point is scheme dependent.
We observe that the fixed point moves to stronger coupling as the HYP smearing parameters in the RG blocking transformation increase.

When we block our $8^4$, $16^4$ and $32^4$ lattices down to a final volume $V_f = 2^4$ (corresponding to $n_b = 4$), the observables become very noisy, making matching more difficult.
The problem grows worse as the HYP smearing parameters increase, and our current statistics do not allow reliable three-lattice matching for $V_f = 2^4$ in schemes~2 and 3.
To resolve this issue, we are accumulating more statistics in existing $32^4$ runs, and generating additional $32^4$ ensembles at more values of the gauge coupling $\be_F$.
These additional data will also improve our results for scheme~1, which we show in \fig{fig:scheme1}.
Different volumes and $n_b$ do not produce identical results in scheme~1, suggesting that the corresponding systematic effects are still non-negligible.
We can estimate finite-volume effects by comparing $n_b = 3$ with $V_f = 3^4$ and $V_f = 4^4$.
Systematic effects due to $n_b$ can be estimated from $n_b = 4$ and $V_f = 2^4$, but this is difficult due to the noise in the $2^4$ data.
Even treating the spread in the results shown in \fig{fig:scheme1} as a systematic uncertainty, we still obtain a clear zero in the bare step-scaling function, indicating an IR fixed point.

% ------------------------------------------------------------------

% ------------------------------------------------------------------
\section{Conclusion} % Draft complete
In this proceedings we have shown how the Wilson-flow-optimized MCRG two-lattice matching procedure proposed in \refcite{Petropoulos:2012mg} improves upon traditional lattice renormalization group techniques.
By optimizing the flow time for a fixed RG blocking transformation, WMCRG predicts a bare step-scaling function $s_b$ that corresponds to a unique discrete \be function.
Applying WMCRG to new 12-flavor ensembles generated with exactly massless fermions, we observe an infrared fixed point in $s_b$.
The fixed point is present for all investigated lattice volumes, number of blocking steps and renormalization schemes, even after accounting for systematic effects indicated by \fig{fig:scheme1}.
This result reinforces the IR-conformal interpretation of our complementary $N_f = 12$ studies of phase transitions~\cite{Schaich:2012fr, Hasenfratz:2013uha}, the Dirac eigenmode number~\cite{Cheng:2013eu, Cheng:2013bca}, and finite-size scaling~\cite{Hasenfratz:2013eka}.
% Something about accumulating more data?  Or would that be less of a high note to go out on?
% ------------------------------------------------------------------

% ------------------------------------------------------------------
\section*{Acknowledgments} % Draft complete
This research was partially supported by the U.S.~Department of Energy (DOE) through Grant No.~DE-SC0010005 (A.~C., A.~H.\ and D.~S.) and by the DOE Office of Science Graduate Fellowship Program under Contract No.~DE-AC05-06OR23100 (G.~P.).
A.~H.\ is grateful for the hospitality of the Brookhaven National Laboratory HET group during her extended visit.
Our code is based in part on the MILC Collaboration's lattice gauge theory software.\footnote{\href{http://www.physics.utah.edu/~detar/milc/}{http://www.physics.utah.edu/$\sim$detar/milc/}}
Numerical calculations were carried out on the HEP-TH and Janus clusters at the University of Colorado, the latter supported by the U.S.~National Science Foundation (NSF) through Grant No.~CNS-0821794; at Fermilab under the auspices of USQCD supported by the DOE; and at the San Diego Computing Center and Texas Advanced Computing Center through XSEDE supported by NSF Grant No.~OCI-1053575.
% ------------------------------------------------------------------

% ------------------------------------------------------------------
\bibliographystyle{utphys}
\bibliography{renorm_pos}
\end{document}